Preprint

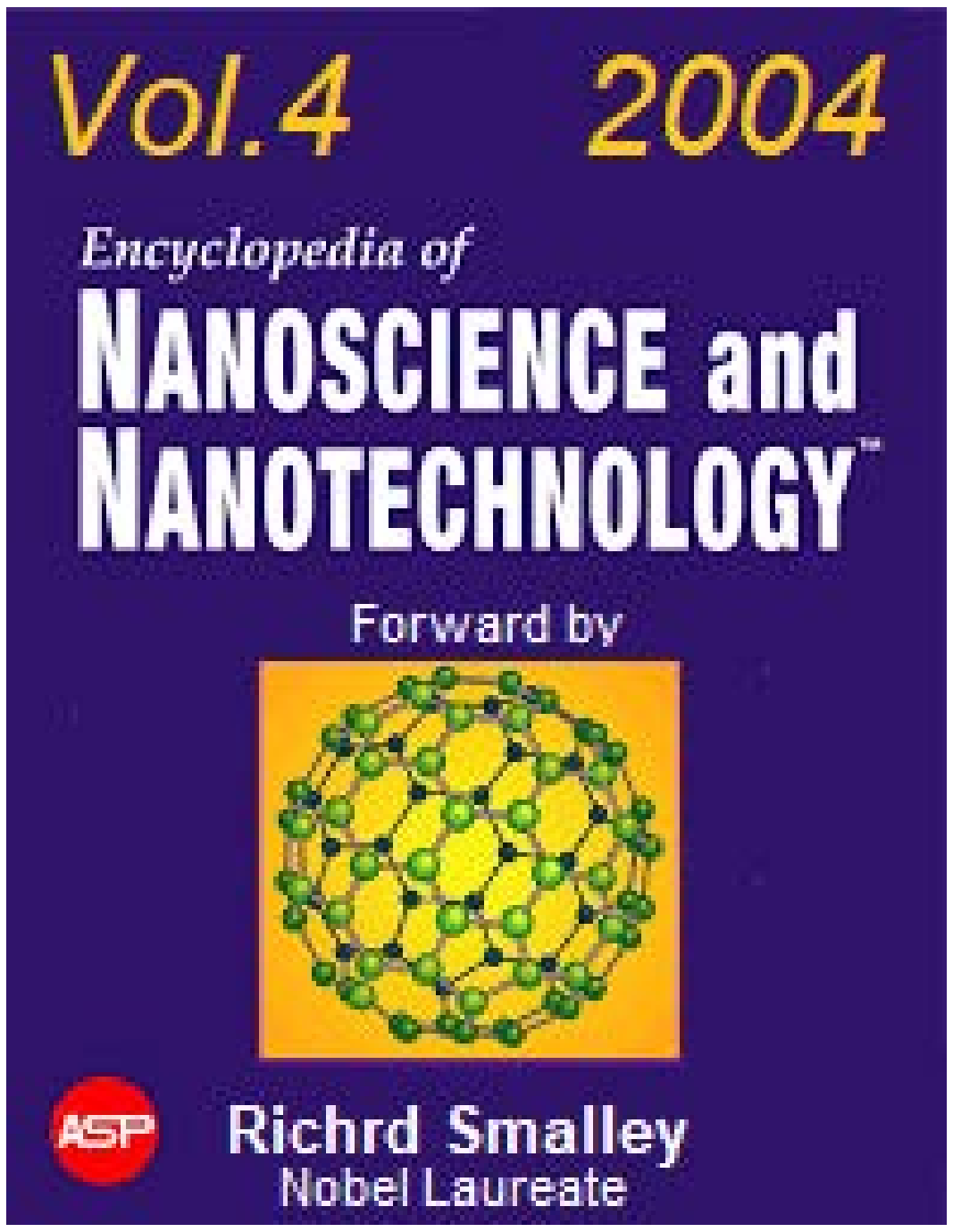

# Interatomic Potential Models for Nanostructures

By

**Hashem Rafii-Tabar** [(1)] and **G.Ali Mansoori** [(2)]

*(1). Institute for Studies in Theoretical Physics and Mathematics (IPM), P.O. Box 19395-5531, Tehran, Iran Email: rafii@theory.ipm.ac.ir*

*(2). University of Illinois at Chicago (M/C 063), Chicago, IL 60607-7052 USA Email: mansoori@uic.edu*

# Table of Contents (TOC)

**Subject** **Page No.**

# 1. Introduction

Over the last decade, nanoscience and nanotechnology [1-4] have emerged as two of the pillars of the research that will lead us to the next industrial revolution [5], and together with molecular biology and information technology, will map the course of scientific and technological developments in the 21st century. This progress has been largely due to the development of sophisticated theoretical and experimental techniques, and practical tools, for understanding, characterizing and manipulating nanoscale structures, processes and systems. On the experimental front, the most significant developments were brought about by the invention of the scanning tunneling microscope (STM) in 1982 [6], followed by the atomic force microscope (AFM) [7] in 1986. These are tip-based devices which allow for a nanoscale manipulation of the morphology of the condensed phases and the determination of their electronic structures. These probe-based techniques have been extended further and are now collectively referred to as the scanning probe microscopy (SPM). The SPM-based techniques have been improved considerably, providing new tools in research in such fields of nanotechnology as nanomechanics, nanoelectronics, nanomagnetism and nanooptics [8].

The fundamental entities of interest to nanoscience and nanotechnology are the isolated individual nanostructures and their assemblies. Nanostructures are constructed from countable (limited) number of atoms or molecules. Their sizes are larger than individual molecules and smaller than micro-structures. Nanoscale is a magical point on the dimensional scale: Structures in nanoscale (called *nanostructures*) are considered at the borderline of the smallest of human-made devices and the largest molecules of living systems. One of their characteristic features is their high surface-to-volume ratio. Their electronic and magnetic properties are often distinguished by quantum mechanical behavior, while their mechanical and thermal properties can be understood within the framework of classical statistical mechanics. Nanostructures can appear in all forms of condensed matter, be it soft or hard, organic or inorganic and/or biological. They form the building blocks of nanotechnology, and the formation of their assemblies requires a deep understanding of the interactions between individual atoms and molecules forming the nanostructures. Accordingly, nanotechnology has been specialized into three broad areas, namely the wet, the dry and the computational nanotechnology.

The wet nanotechnology is mainly concerned with the study of nanostructures and nanoprocesses in biological and organic systems that exist in aqueous environment. An important aspect of research in wet nanotechnology is the design of smart drugs for targeted delivery using such nanostructures as nanotubes and self-assembling materials [9, 10] as platforms. The dry nanotechnology, on the other hand, addresses electronic and mechanical properties of metals, ceramics, focusing on fabrication of structures in carbon (e.g. fullerenes and nanotubes), silicon, and other inorganic materials.

The computational nanotechnology is based on the fields of mathematical modeling and computer-based simulation [11] that allow for computation and prediction of the underlying dynamics of nanostructures and processes in condensed matter physics, chemistry, materials science, biology and genetics. Computational nanotechnology, therefore, covers the other domains of nanofields by employing concepts from both classical and quantum mechanical many body theories. It can provide deep insights into the formation, evolution and properties of nanostructures and mechanisms of nanoprocesses. This is achieved by performing precise atom-by-atom numerical experiments (modeling and simulation) on many aspects of various condensed phases. The precision of such calculations depends on the accuracy of the interatomic and intermolecular potential energy functions at hand.

At the nanoscale, the implementation of the computational science leads to the study of the evolution of physical, chemical and biophysical systems on significantly reduced length, time and energy scales. Computer simulations at this scale form the basis of computational nanoscience. These simulations could allow for an understanding of the atomic and molecular scale structures, energetics, dynamics and mechanisms underlying the physical and chemical processes that can unfold in isolated nanostructures, and their assemblies, under different ambient conditions.

This review is concerned with one of the most important elements of the computational approach to the properties of, and processes involving, nanoscale structures, namely the phenomenological interatomic and intermolecular potentials. The mathematical expressions for the phenomenological forces and potential energies between atoms and molecules necessary for prediction of bulk (macroscopic) fluid and solid properties are rather well understood [12-14]. There are sufficient, effective phenomenological intermolecular potential energy functions available for the statistical mechanics prediction of macroscopic systems [13-17]. Parameters of phenomenological interaction energies between atoms and simple molecules can be calculated through such measurements as x-ray crystallography, light scattering, nuclear magnetic resonance spectroscopy, gas viscosity, thermal conductivity, diffusivity and the virial coefficients data [18]. Most of the present phenomenological models for interparticle forces are tuned specifically for statistical mechanical treatment of macroscopic systems. However, such information may not be sufficiently accurate in the treatment of nanosystems where the number of particles are finite and the statistical averaging techniques fail.

Nanostructures consist of many body systems, and a rigorous modeling of their properties has to be placed within the quantum mechanical domain, taking into account the electronic degrees of freedom. For simple atoms and molecules the quantum mechanical *ab initio* calculation methods [19] have been successful to produce accurate intermolecular potential functions. While *ab initio* calculations may be satisfactory for simple molecules, for complex molecules and macromolecules they may not be able to produce the accurate needed information. However, even with today's enhanced computational platforms and sophisticated quantum mechanical techniques [20], the nanostructures that can be studied from a quantum mechanical, or *ab initio*, basis are those composed of at most a few hundred atoms. Consequently, the use of phenomenological interatomic and intermolecular potentials in simulations is still necessary. This allows modeling of nanostructures consisting of several millions of atoms, and recently simulations involving more than $10^9$ atoms have been performed

To motivate the use of interatomic and intermolecular potentials and show how they enter into nanoscale modeling, we consider, in Section 2, one of the widely used methods for numerical modeling at the nanoscale. This is followed, in Section 3, by a description of several types of state-of-the-art interatomic potentials that are in current use for modeling the energetics and dynamics of several classes of materials, including, metals, semi-metals and semi-conductors. We will then briefly review the applications of these potentials in specific computational modeling studies.

## 2. Computer-Based Simulation Methods

Computer simulations applied in nanoscience consist of computational "experimentations" conducted on an assembly of countable number of molecules with the assumption of predefined intermolecular interaction models. Computer simulations can direct an experimental procedure and have the potential of replacing an experiment if accurate intermolecular potentials are used in their development.

Computer simulation modeling of the physics and chemistry of nanostructures composed of several millions to several hundreds of millions of atoms, can be performed by employing several distinct approaches. The most widely used approaches include (1) Monte Carlo simulation, (2) Molecular Dynamics simulation.

The cell in which the simulation is performed is replicated in all spatial dimensions, generating its own periodic images containing the periodic images of the original N atoms. This is the periodic boundary condition, and is introduced to remove the undesirable effects of the artificial surfaces associated with the finite size of the simulated system. The forces experienced by the atoms and molecules are obtained from prescribed two-body or many-body interatomic and intermolecular potentials, $H_l(r_{ij})$, according to

$$F_i = - \sum_{j>i} \nabla_{\mathbf{r_i}} H_l(r_{ij}) \ , \qquad (1)$$

where $r_{ij}$ is the separation distance between two particles i and j.

### 2.1. Monte Carlo (MC) Simulation Methods

The Metropolis Monte Carlo (MC) simulation methods can be used in nanoscience to simulate various complex physical phenomena including prediction of phase transitions, thermally-averaged structures and charge distributions, just to name a few [21]. There exist variety types of MC simulations which are used depending on the nano system under consideration and the kind of computational results in mind. They include, but not limited to, Classical MC, Quantum MC and Volumetric MC. In the Classical MC the classical Boltzmann distribution is used as the starting point to perform various property calculations. Through the use of Quantum MC one can compute quantum-mechanical energies, wave functions and electronic structure using Schroedinger's equation. The Volumetric MC is used to calculate molecular volumes and sample molecular phase-space surfaces [22].

## 2.2. Molecular Dynamics (MD) Simulation Method

In the Molecular Dynamics (MD) simulation methods [23-25] the emphasis is on the motion of individual atoms within an assembly of N atoms, or molecules, that make up the nanostructure under study. The dynamical theory employed to derive the equations of motion is either the Newtonian deterministic dynamics or the Langevin-type stochastic dynamics. The initial data required are the initial position coordinates and velocities of the particles, in either a crystalline or an amorphous state, located in a primary computational cell of volume V .

To save computational time, the simplifying assumption is made that each particle interacts with its nearest neighbors, located in its own cell as well as in the image cells, that are within a specified cut-off radius. The 3N coupled differential equations of motion can then be solved by a variety of numerical finite-difference techniques, one of which is the velocity Verlet algorithm [23], according to which the positions, $r_i$, and velocities, $v_i$, of the particles of mass $m_i$ are updated at each time step, $dt$, by

$$r_i(t + dt) = r_i(t) + v_i(t)dt + (½)\, dt^2\, F_i(t)/m_i\ ,$$

$$v_i(t + dt/2) = v_i(t) + (½)\, dt\, F_i(t)/m_i, \qquad (2)$$

$$v_i(t + dt) = v_i(t + dt/2) + (½)dt\, F_i(t + dt)/m_i.$$

The dynamical history of a particular micro-state of the system, constructed initially, is followed by computing the space-time trajectories through the phase space via Eqs. (2). At each instant of the simulation time, the exact instantaneous values of the observables, such as pressure, temperature and thermodynamics response functions, are also obtained, leading to time-average values at the conclusion of the simulation.

### 2.2.1. Constant Temperature MD Simulation: Nosé -Hoover dynamics

For a large class of problems in the physics and chemistry of nanostructures, the type of system that is considered is a closed one. This is a system with a fixed volume, V, a fixed number of particles, N, maintained at a constant temperature, T. Within statistical mechanics, such a system is represented by a constant (NVT), or canonical, ensemble [26], where the temperature acts as a control parameter.

A constant-temperature MD simulation can be realized in a variety of ways. A method that generates the canonical ensemble distribution in both the configuration space and momentum space parts of the phase space was proposed by Nosé [27-29] and Hoover [30] and is referred to as the extended system method. According to this method, the simulated system and a heat bath couple to form a composite system. This coupling breaks the energy conservation that otherwise restricts the behavior of the simulated system and leads to the generation of a canonical ensemble. The conservation of energy still holds in the composite system, but the total energy of the simulated system is allowed to fluctuate.

The mathematical formulation of the method is based on the extension of the space of dynamical variables of the system beyond that of the coordinates and momenta of the real particles to include one additional *phantom coordinate*, $s$, and its conjugate momentum, $p_s$, [31]. This extra degree of freedom acts as a heat bath for the real particles. There are, therefore, four systems to consider, namely, the real ($\vec{r}_i$, $\vec{p}_i$) system, the virtual ($\tilde{r}_i$, $\tilde{p}_i$) system, the real extended ($\vec{r}_i$, $\vec{p}_i$, $s$, $p_s$) system and the virtual extended ($\tilde{r}_i$, $\tilde{p}_i$, $s$, $p_s$ ) system. The aim of the Nosé's approach is to show that there is a method for selecting the Hamiltonian of the extended system and, simultaneously, to relate the variables of the real system to those of the virtual system, such that the micro-canonical partition function of the extended virtual system is proportional to the canonical partition function of the real system [31].

The Hamiltonian of the virtual extended system is

$$H^* = \sum_{i\to N} \left[ \vec{\tilde{p}}_i^{\,2}/(2ms^2) \right] + H_I(\vec{\tilde{r}}_{ij}) + p_s^2/(2Q) + g\, k_B T \ln s \, , \qquad (3)$$

where g is the number of degrees of freedom, $k_B$ is the Boltzmann constant, Q is a parameter which behaves like a 'mass' associated with the motion of the coordinate s, and $\vec{r}_i$, $\vec{p}_i$ and $\vec{\tilde{r}}_i$, $\vec{\tilde{p}}_i$ are the canonical position and momentum coordinates of all the particles in the real and virtual systems, respectively. The virtual coordinates, and the time, are related to the corresponding real coordinates via the transformations

$$\begin{aligned} \vec{r}_i &= \vec{\tilde{r}}_i \\ \vec{p}_i &= (1/s)\, \vec{\tilde{p}}_i \\ dt &= (1/s)\, d\tilde{t} \end{aligned} \qquad (4)$$

Since $H_I$ in Eq. (3) is the potential energy for both the real and virtual systems, then the first two terms in the right hand side of Eq. (3) represent the kinetic and potential energies of the real system, respectively, and the last two terms correspond to the kinetic and potential energies, respectively, associated with the extra degree of freedom.

From this Hamiltonian the equations of motion of the real system are obtained

$$\begin{aligned} dr_i/dt &= \vec{p}_i \,/\, m_i \, , \\ dp_i/dt &= \mathbf{F}_i - \eta\, \vec{p}_i \, , \\ d\eta/dt &= (1/Q) \left[ \sum_i ( \vec{p}_i^{\,2} / m_i) - g k_B T \right], \end{aligned} \qquad (5)$$

where $\eta$ is called the friction coefficient of the bath. This coefficient is not a constant and can take on both positive and negative values. This gives rise to what is called a negative feedback mechanism. The last equation in (5) controls the functioning of the heat bath. From this equation we observe that if the total kinetic energy is greater than $gk_BT/2$ then $d\eta/dt$, and hence $\eta$, is positive. This prompts a friction inside the bath and correspondingly the motion of the atoms are decelerated to lower their kinetic energy to that of the bath. On the other hand, if the kinetic energy is lower than $gk_BT/2$, then $d\eta/dt$ will be negative, and this results in the bath being heated up and accelerate the motion of the atoms. Equations (5) are collectively referred to as the Nosé-Hoover thermostat.

### 2.2.2. Equations of motion

The implementation of the Nosé-Hoover dynamics substantially modifies Eq.s (2), the equations of motion. A velocity Verlet version of this dynamics formulation can be given by the following expressions [32]

$$r_i(t + dt) = r_i(t) + v_i(t)dt + (½)dt^2[F_i(t)/ m_i - \eta(t)v_i(t)],$$

$$v_i(t + dt/2 ) = v_i(t) + (dt/2) [ F_i(t)/ m_i - \eta(t)v_i(t)],$$

$$\eta(t + dt/2 ) = \eta(t) + [dt/(2Q)] \Big[\sum_{i\to N} m_i v_i^2(t) - gk_B T\Big], \qquad (6)$$

$$\eta(t + dt) = \eta(t + dt/2) + [dt/(2Q)] \Big[\sum_{i\to N} m_i v_i^2(t + dt/2 ) - gk_BT\Big],$$

$$v_i(t + dt) = 2 [v_i(t + dt/2) + dt F_i(t + dt)/(2m_i)] / [2 + \eta(t + dt)dt].$$

A particular parameterization of Q is given by

$$Q = g k_BT \tau^2, \qquad (7)$$

where $\tau$ is the relaxation time of the heat bath, normally of the same order of magnitude as the simulation time step, *dt*. It controls the speed with which the bath damps down the fluctuations in the temperature. The number of degrees of freedom is given by $g = 3(N - 1)$.

## 3. Interatomic potentials

To study nanostructures composed of several hundred to several million atoms or molecules, the computationally most efficient method is the use of phenomenological interatomic and intermolecular potentials. This is because the existing quantum mechanical techniques are able to deal with at most a few hundred atoms.

The phenomenological potentials are obtained by using phenomenological approaches of selecting a mathematical function and fitting its unknown parameters to various, experimentally determined, properties of the system, such as its lattice constant.

Interatomic and intermolecular potentials must be able to model the energetics and dynamics of nanostructures, and this fact lies at the very foundation of the computer-based modeling and simulations. Potentials describe the physics of the model systems, and the significance of much of the modeling and simulation results, their accuracy and the extent to which they represent the real behavior of nanostructures, and their transitions, under varied conditions, depends in a critical manner on the accuracy of the interatomic and intermolecular potentials employed.

A great deal of effort has been spent over the years to develop phenomenological intermolecular potentials to model the bonding in various classes of materials, such as metallic, semi-metallic, semi-conducting, and organic atoms and molecules. For a review see [11,33,34].

Basically intermolecular potential energies include pairwise additive energies, as well as many body interactions.

The interparticle interaction potential energy between atoms and molecules is generally denoted by $H(r)=H_{rep}+H_{att}$ where r is the intermolecular distance, $H_{rep}$ is the repulsive interaction energy and $H_{att}$ is the attractive interaction energy, see Figure 1.

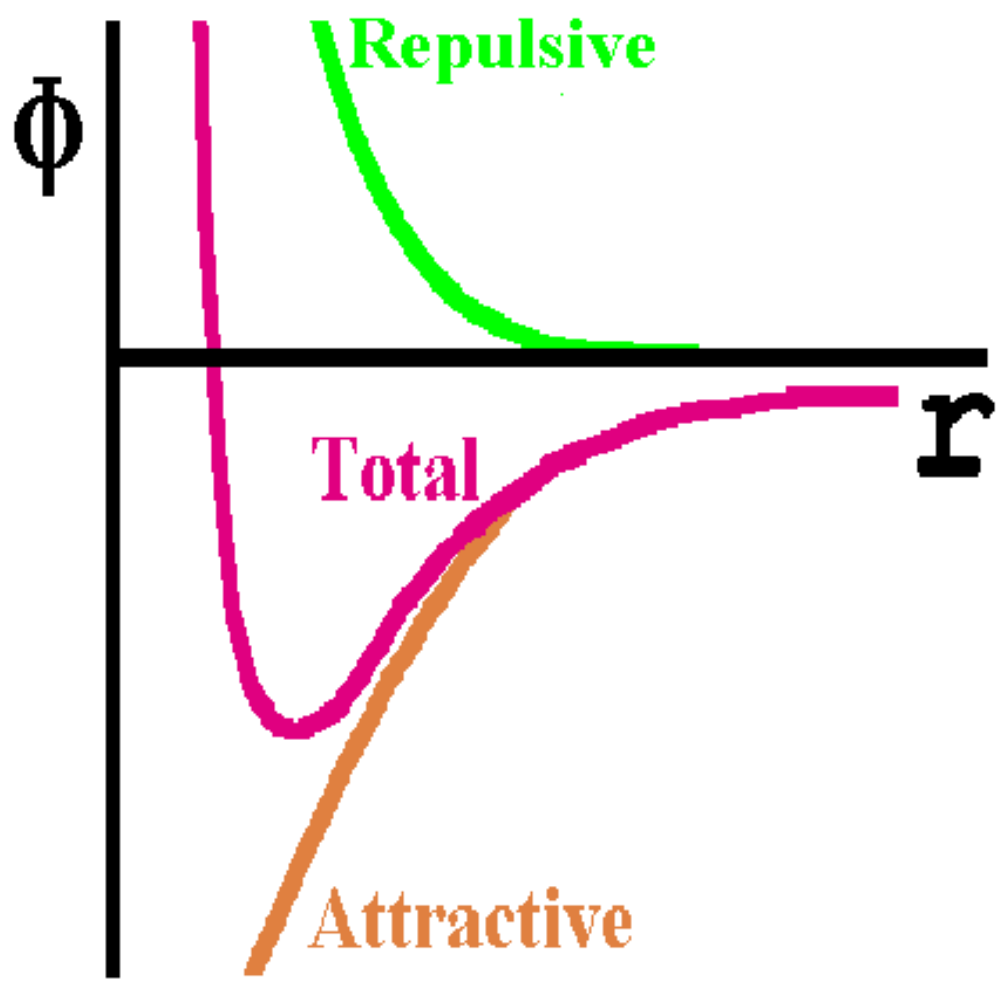

***Figure 1**: The pair interaction energy*

From the equation above, the interaction force is

$$\mathbf{F} = -\nabla H(\mathbf{r}) = \mathbf{F}_{rep} + \mathbf{F}_{att}$$

For neutral and spherically symmetric molecules when the separation (r) is very small, an exponential repulsive term, $H_{rep}=\alpha exp(-\beta r)$, dominates, and the potential is strongly positive. Hence the $H_{rep}=\alpha exp(-\beta r)$ term describes the short-range repulsive potential due to the distortion of the electron clouds at small separations. For neutral and spherically symmetric molecules when the separation (r) is large the London dispersion forces dominate.

Among pairwise additive energies one can mention the repulsive potentials, van der Waals energies, interactions involving polar and polarization of molecules, interactions involving hydrogen bonding and strong intermolecular energies including covalent and coulomb interactions [35, 36]. Among many body interactions one can name the Axilrod-Teller triple-dipole interactions [37-39].

To be effective for computational nanotechnology, interatomic and intermolecular potentials must possess the following properties [40,41]:

a) Flexibility: A potential energy function must be sufficiently flexible that it could accommodate as wide a range as possible of fitting data. For solid systems, this data might include lattice constants, cohesive energies, elastic properties, vacancy formation energies, and surface energies.

b) Accuracy: A potential function should be able to accurately reproduce an appropriate fitting database.

c) Transferability: A potential function should be able to describe at least qualitatively, if not with quantitative accuracy, structures not included in a fitting database.

d) Computational efficiency: Evaluation of the function should be relatively efficient depending on quantities such as system sizes and time-scales of interest, as well as available computing resources.

In this section we shall describe some of the potential functions that meet these criteria, and are widely used in computational nanoscience.

### 3.1. Interatomic potentials for metallic systems

Bonding in metallic systems operates over the range of 0.2 to 0.5 nm [42]. At large interatomic distances, the predominant forces arise from van der Waals interactions, which are responsible for long-range cohesion. Metallic bonding, like covalent bonding, arises from the sharing of electrons and hence its proper description requires the consideration of the many-body effects. Two-body potentials are incapable of describing this bonding [43, 44] since:

a): For most cubic metals, the ratio of the elastic constants, $C_{12}$ to $C_{44}$, is far from unity, whereas a pairwise potential leads to the Cauchy relation, i.e. $C_{12} = C_{44}$.

b): The prediction of the unrelaxed vacancy formation energy gives values around the cohesive energy which is completely incorrect for metals. The relaxation energy for metals is quite small and the experimental data suggest that the vacancy formation energy for metals is about one third of the cohesive energy.

c): The interatomic distance between the first and second atomic layers within an unreconstructed surface structure (bulk cross section) is predicted to be expanded by pairwise potentials. This is in contrast with the experimental data which suggest a contraction of the open surface lattice spacing. i.e. pair potentials fail to predict an inward relaxation of the metallic surfaces.

d): Pairwise potentials overestimate the melting point by up to 20% of the experimental value.

e): Potentials with a functional form having only one optimum at the diatomic equilibrium distance cannot be fitted properly to the phonon frequencies.

Two approaches have been proposed for going beyond pair potentials and incorporating many-body effects into two-body potentials:

The first approach is to add a term, which is a functional of the local electronic density of a given atom, to the pairwise term. This method has itself led to several alternative potentials that mimic the many-body effects. These many-body potentials are known as the embedded-atom model (EAM) potentials [45-47], which have been employed in several studies involving elemental metals and their alloys [48-53], the Glue Model potentials [54], the Finnis-Sinclair potentials for the BCC elemental metals [55], which have also been developed for the noble metals [56], the Sutton-Chen (SC) potentials [57] for the ten FCC elemental metals, and the Rafii-Tabar and Sutton potentials [58] for the FCC random binary alloys which have also been used in several modeling studies [11, 59-61].

The second approach is to go from pair potentials to cluster potentials by the addition of higher order interactions, for example three-body and four-body terms, with appropriate functional forms and symmetries. This has led to potentials, such as the Murrell-Mottram cluster potentials [44]. Inclusion of higher-order terms provides a more accurate modeling of the energetics of the phenomena than is given by pair potentials alone. In the following sections, we consider the potentials pertinent to each approach.

### 3.1.1. The many-body embedded-atom model (EAM) potentials

The many-body EAM potentials were proposed [45-47] to model the bonding in metallic clusters. They were the first alternatives to the traditional pair potential models.

Their construction is based on the use of density functional theory (DFT), according to which the energy of a collection of atoms can be expressed exactly by a functional of its electronic density [62]. Similarly, the energy change associated with embedding an atom into a host background of atoms is a functional of the electronic density of the host before the new atom is embedded [63, 64]. If we can find a good approximation to the embedding functional, then an approximate expression for the energy of an atom in a metal can be constructed.

The total electron density of the host atoms is approximated as a linear superposition of the electron densities (charge distributions) of individual host atoms. To zeroth order, the embedding energy can be equated to the energy of embedding an atom in a homogenous electron gas, whose density, $\rho_{h,i}$ , matches the host density at the position of the embedded atom, augmented by the classical electrostatic interaction with the atoms in the host system [65]. The embedding energy for the homogeneous electron gas can be calculated from an *ab initio* basis. Computation of $\rho_{h,i}$ from a weighted average of the host density over the spatial extent of the embedded atom improves the description by accounting for the local inhomogeneity of the host density. The classical electrostatic interaction reduces to a pairwise sum if a frozen atomic charge density is assumed for each host atom [65]. This approach, called quasi-atom method [63], or the effective-medium theory [64], provides the theoretical basis of the EAM,
and similar methods.

In the EAM model, the total energy of an elemental system is, therefore, written as

$$H_l^{EAM} = \sum_i F_i [\rho_{h,i}] + (½)\sum_i \sum_{j \neq i} \phi_{ij} (r_{ij}) , \qquad (8)$$

where $\rho_{h,i}$ is electron density of the host at the site of atom $i$, $F_i[\rho]$ is the embedding functional, *i.e.* the energy to embed the atom $i$ into the background electron density, $\rho$, and $\phi_{ij}$ is a pairwise central potential between atoms $i$ and $j$, separated by a distance $r_{ij}$, and represents the repulsive core-core electrostatic interaction. The host electron density is a linear superposition of the individual contributions, and is given by

$$\rho_{h,i} = \sum_{j \neq i} \rho_j^*(r_{ij}) , \qquad (9)$$

where $\rho_j^*$, another pairwise term, is the electron density of atom $j$ as a function of interatomic separation. It is important to note that the embedding functional, $F_i[\rho]$, is a universal functional that does not depend on the source of the background electron density. This implies that the same functional is employed to compute the energy of an atom in an alloy as that employed for the same atom in a pure elemental metal [48]. Indeed, this is one of the attractive features of these potentials. For a solid at equilibrium, the force to expand, or contract, due to the embedding function is exactly balanced by the force to contract, or expand due to the pairwise interactions. At a defect, this balance is disrupted, leading to the displacements as atoms move to find a new balance [65]. The positive curvature of $F$ plays a key role in this process, by defining the optimum trade off between the number of bonds and the length of those bonds.

The expression for the Cauchy pressure for a cubic crystal can be found from Eq. (8), and is seen to depend directly on the curvature of the function *F as described in* [46]

$$C_{11} - C_{44} = (1/\Omega)(d^2F / d\rho_{h,i}^2)[\sum_j (d\rho/dr_{ij})(x_{ij}^2/r_{ij})]^2, \qquad (10)$$

where $\Omega$ is the atomic volume and $x_{ij}$ is the x-component of the $r_{ij}$.

To apply these potentials, the input parameters required are the equilibrium atomic volume, the cohesive energy, the bulk modulus, the lattice structure, as well as the repulsive pair potentials and the electron density function [50]. Among the extensive applications of these potentials, we can list their parameterization and use in the computation of the surface energy and relaxation of various crystal surfaces of *Ni* and *Pd* and the migration of hydrogen impurity in the bulk *Ni* and *Pd* [46], the computation of the formation energy, migration energy of vacancies and surface energies of a variety of FCC metals [48], the calculation of the surface composition of the *Ni-Cu* alloys [66], the computation of the elastic constants and vibrational modes of the $Ni_3Al$ alloy [49], the self-diffusion and impurity diffusion of the FCC metals [51], the computation of the heats of solution for alloys of a set of FCC metals [52], and the computation of the phase stability of FCC alloys [53]. There has also been an application of these potentials to covalent materials, such as *Si* [67].

In a recent application [68], the second-order elastic moduli ($C_{11}$, $C_{12}$, $C_{44}$) and the third-order elastic moduli ($C_{111}$, $C_{112}$, $C_{123}$, $C_{144}$, $C_{166}$, $C_{456}$), as well as the cohesive energies and lattice constants, of a set of 12 cubic metals with FCC and BCC structures were used as input to obtain the corresponding potential parameters for these metals [69]. The resulting potentials were then used to compute the pressure-volume (P - V ) curves, phase stabilities and the phonon frequency spectra, with excellent agreement obtained for the P - V curves with the experimental data, and a reasonable agreement obtained for the frequency curves.

The EAM potentials can also be written for ordered binary alloys [65]. We can write

$$H_{Alloy}^{EAM} = \sum_i F_{ti}\,[\rho_{h,i}] + (½)\sum_i\sum_{j\neq i} \phi_{ti,tj}\,(r_{ij}), \qquad (11)$$

where $\phi$ now depends on the type of atom $t_i$ and atom $t_j$. The host electron density is now given by

$$\rho_{h,i} = \sum_{j\neq i} \rho^*_{tj}\,(r_{ij}), \qquad (12)$$

where the terms in the sum each depends on the type of neighbor atom *j*. Therefore, for a binary alloy with atom types *A* and *B*, the EAM energy requires definitions for $\phi_{AA}(r)$, $\phi_{BB}(r)$, $\phi_{AB}(r)$ , $\rho_A(r)$, $\rho_B(r)$, $F_A(\rho)$ and $F_B(\rho)$.

### 3.1.2. The many-body Finnis and Sinclair (FS) potentials

These potentials [55] were initially constructed to model the energetics of the transition metals. They avoid the problems associated with using pair potentials to model metals, e.g. the appearance of the Cauchy relation between the elastic constants $C_{12}$=$C_{44}$ which is not satisfied by cubic crystals. They also offer a better description of the surface relaxation in metals.

In the FS model, the total energy of an N-atom system is written as

$$H_I^{FS} = (½)\sum_{i\to N}\sum_{j\neq i}V(r_{ij}) - c\sum_i (\rho_i)^{1/2} , \qquad (13)$$

Where

$$\rho_i = \sum_{j\neq i} \phi( r_{ij}) \ . \qquad (14)$$

The function $V(r_{ij})$ is a pairwise repulsive interaction between atoms i and j, separated by a distance $r_{ij}$, $\phi(r_{ij})$ are two-body cohesive pair potentials and $c$ is a positive constant. The second term in Eq. (13) represents the cohesive many-body contribution to the energy. The square root form of this term was motivated by an analogy with the second moment approximation to the Tight-Binding Model [70]. To see this, we start with the tight-binding approach [71] in which the total electronic band energy, *i.e.* the total bonding energy, which is given as the sum of the energies of the occupied one-electron states, is expressed by

$$E_{tot} = 2\int_{-\infty}^{E_f} E.n(E).dE , \qquad (15)$$

where $n(E)$ is the electron density of states, $E_f$ is the Fermi level energy and the factor 2 refers to spin degeneracy. $E_{tot}$ is an attractive contribution to the configurational energy, which is dominated by the broadening of the partly filled valence shells of the atoms into bands when the solid is formed [72]. It is convenient to divide $E_{tot}$ into contributions from individual atoms

$$E_{tot} = \sum_i E_i = 2 \sum_i \int_{-\infty}^{E_f} E.n_i(E)dE , \qquad (16)$$

$$n_i(E) = \sum_{\upsilon}|< \Psi_{\upsilon} \mid i >|^2 \ \delta (E - E_{\upsilon}), \qquad (17)$$

is the projected density of states on site *i* and $|\Psi_\upsilon>$ are the eigenfunctions of the one-electron Hamiltonian. As has been discussed in [72]. To obtain $n_i(E)$ exactly, it is in principle necessary to know the positions of all atoms in the crystal. Furthermore, $n_i(E)$ is a very complicated functional of these positions. However, it is not necessary to calculate the detailed structure of $n_i(E)$. To obtain an approximate value of quantities such as $E_i$ which involves integrals over $n_i(E)$, we need only information about its width and gross features of its shape. This information is conveniently summarized in the moments of $n_i(E)$, defined by

$$\mu_n^{\,i} = \int_{-\infty}^{\infty} E^n n_i(E)dE \tag{18}$$

The important observation, which allows a simple description comparable to that of interatomic potentials, is that these moments are rigorously determined by the local environment. The exact relations are [72]

$$\mu^i_2 = \sum_j h_{ij}^2$$
$$\mu^i_3 = \sum_{jk} h_{ij}h_{jk}h_{ki}$$
$$\mu^i_4 = \sum_{jkl} h_{ij}h_{jk}h_{ki}h_{li}\ , \tag{19}$$

where

$$h_{ij} =< \chi_i \mid H \mid \chi_j >, \tag{20}$$

and $\chi_i$ is the localized orbital centered on atom *i*, and *H* is the one-electron Hamiltonian. Therefore, if we have an approximate expression for the $E_i$ in terms of the first few $\mu^i_n$, the electronic band energy can be calculated with essentially the same machinery used to evaluate interatomic potentials. Now, the exact evaluation of $E_i$ requires the values of all the moments on site *i*. However, a great deal of information can be gained from a description based only on the second moment, $\mu^i_2$. This moment provides a measure of the squared valence-band width, and thus sets a basic energy scale for the problem. Therefore, a description using only $\mu^i_2$ assumes that the effects of the structure of $n_i(E)$ can be safely ignored, since the higher moments describe the band shape. Since, $E_i$ has units of energy and $\mu^i_2$ has units of (energy)$^2$ , therefore we have

$$E_i = E_i(\mu^i_2) = - A \sqrt{(\mu^i_2)} = - A \sqrt{(\sum_j h_{ij}^2)} \; , \qquad (21)$$

where $A$ is a positive constant that depends on the chosen density of states shape and the fractional electron occupation [65].

The functions $\phi(r_{ij})$ in Eq. (14) can be interpreted as the sum of squares of hopping (overlap) integrals. The function $\rho_i$ can be interpreted as the local electronic charge density [45] constructed by a rigid superposition of the atomic charge densities $\phi(r_{ij})$. In this interpretation, the energy of an atom at the site $i$ is assumed to be identical to its energy within a uniform electron gas of that density. Alternatively, $\rho_i$, can be interpreted [55] as a measure of the local density of atomic sites, in which case Eq. (13) can be considered as a sum consisting of a part that is a function of the local volume, represented by the second term, and a pairwise interaction part, represented by the first term. The FS potentials, Eq. (13), are similar in form to the EAM potentials in Eq. (8). However, their interpretations are quite different. The FS potentials, as has been shown above, were derived on the basis of the Tight-Binding Model and this is the reason why their many-body parts, which correspond to the $F_i\,[\rho_{h,i}]$ functionals in the EAM potentials, are in the form of square root terms. Furthermore, the FS potentials are less convenient than the EAM potentials for a conversion from the pure metals to their alloys. Notwithstanding this difficulty, FS potentials have been constructed for several alloy systems, such as the alloys of the noble metals (Au, Ag, Cu) [56].

### 3.1.3. The many-body Sutton and Chen (SC) long-range potentials

The SC potentials [57] describe the energetics of ten FCC elemental metals. They are of the FS type and therefore similar in form to the EAM potentials. They were specifically designed for use in computer simulations of nanostructures involving a large number of atoms.

In the SC potentials, the total energy, written in analogy with Eq. (13), is given by

$$H_I^{SC} = \varepsilon \left[ (1/2) \sum_i \sum_{j \neq i} V(r_{ij}) - c \sum_i (\rho_i)^{1/2} \right], \tag{22}$$

Where

$$V(r_{ij}) = (a/r_{ij})^n \tag{23}$$

and

$$\rho_i = \sum_{j \neq i} (a/r_{ij})^m, \tag{24}$$

where $\varepsilon$ is a parameter with the dimensions of energy, $a$ is a parameter with the dimensions of length and is normally taken to be the equilibrium lattice constant, $m$ and $n$ are positive integers with $n > m$. The power-law form of the potential terms was adopted so as to construct an unified model that can combine the short-range interactions, afforded by the N-body second term in Eq. (22) and useful for the description of surface relaxation phenomena, with a van der Waals tail that gives a better description of the interactions at the long range. For a particular FCC elemental metal, the potential in Eq. (22) is completely specified by the values of $m$ and $n$, since the equilibrium lattice condition fixes the value of $c$. The values of the potential parameters, computed for a cut-off radius of 10 lattice constants, are listed in Table I. These parameters were obtained by fitting the experimental cohesive energies and lattice parameters exactly. The indices $m$ and $n$ were restricted to integer values, such that the product $mxn$ was the nearest integer to $18\Omega^f B^f/E^f$, Eq. (9) in [57], where $\Omega^f$ is the FCC atomic volume, $B^f$ is the computed bulk modulus, and $E^f$ is the fitted cohesive energy.

The SC potentials have been applied to the computation of the elastic constants, bulk moduli and cohesive energies of the FCC metals, and the prediction of the relative stabilities of the FCC, BCC and HCP structures [57]. The results show reasonable agreement with the experimental values. These potentials have also been used in modeling the structural properties of metallic clusters in the size range of 13 to 309 atoms [73].

| **Element** | ***m*** | ***n*** | ***ε (eV)*** | ***c*** |
|---|---|---|---|---|
| Ni | 6 | 9 | $1.5707 \times 10^{-2}$ | 39.432 |
| Cu | 6 | 9 | $1.2382 \times 10^{-2}$ | 39.432 |
| Rh | 6 | 12 | $4.9371 \times 10^{-3}$ | 144.41 |
| Pd | 7 | 12 | $4.1790 \times 10^{-3}$ | 108.27 |
| Ag | 6 | 12 | $2.5415 \times 10^{-3}$ | 144.41 |
| Ir | 6 | 14 | $2.4489 \times 10^{-3}$ | 334.94 |
| Pt | 8 | 10 | $1.9833 \times 10^{-2}$ | 34.408 |
| Au | 8 | 10 | $1.2793 \times 10^{-2}$ | 34.408 |
| Pb | 7 | 10 | $5.5765 \times 10^{-3}$ | 45.778 |
| Al | 6 | 7 | $3.3147 \times 10^{-2}$ | 16.399 |

*Table I: Parameters of the Sutton-Chen potentials.*

### 3.1.4. The many-body Murrell-Mottram (MM) many-body potentials

The Murrell-Mottram potentials are an example of cluster-type potentials, and consist of sums of effective two- and three body interactions [44, 74, 75]

$$U_{tot} = \sum_i \sum_{j>i} U_{ij}^{(2)} + \sum_i \sum_{j>i} \sum_{k>j} U_{ijk}^{(3)}. \quad (25)$$

The pair interaction term is modeled by a Rydberg function which has been used for simple diatomic potentials. In the units of reduced energy and distance, it takes the form

$$U_{ij}^{(2)}/D = -(1 + a_2 \rho_{ij})\exp(-a_2 \rho_{ij}), \quad (26)$$

where

$$\rho_{ij} = (r_{ij} - r_e)/r_e . \quad (27)$$

*D* is the depth of the potential minimum, corresponding to the diatomic dissociation energy at $\rho_{ij}$=0, i.e. for $r_{ij}=r_e$, with $r_e$ the diatomic equilibrium distance. *D* and $r_e$ are fitted to the experimental cohesive energy and lattice parameter respectively. The only parameter involved in the optimization of the potential is $a_2$, which is related to the curvature (force constant) of the potential at its minimum [44, 74, 75]. The three-body term must be symmetric with respect to the permutation of the three atoms indices, *i, j* and *k*. The most convenient way to achieve this is to create functional forms which are combinations of interatomic coordinates, $Q_1$, $Q_2$ and $Q_3$ which are irreducible representations of the S3 permutation group [76]. If we construct a given triangle with atoms (i, j, k), then the coordinates $Q_i$ are given by

$$\begin{bmatrix} Q_1 \\ Q_2 \\ Q_3 \end{bmatrix} = \begin{vmatrix} \sqrt{1/3} & \sqrt{1/3} & \sqrt{1/3} \\ 0 & \sqrt{1/2} & -\sqrt{1/2} \\ \sqrt{2/3} & -\sqrt{1/6} & -\sqrt{1/6} \end{vmatrix} \begin{bmatrix} \rho_{ij} \\ \rho_{jk} \\ \rho_{ki} \end{bmatrix} \tag{28}$$

with

$$\rho_{\alpha\beta} = (r_{\alpha\beta} - r_e)/r_e, \tag{29}$$

and $r_{\alpha\beta}$ represents one of the three triangle edges ($r_{ij}$, $r_{jk}$, $r_{ki}$). These interatomic coordinates have specific geometrical meanings. $Q_1$ represents the perimeter of the triangle in reduced units, $Q_2$ and $Q_3$ measure the distortions from an equilateral geometry [44]. All polynomial forms which are totally symmetric in $\rho_{\alpha\beta}$ can be expressed as sums of products of the so called integrity basis [44], defined as:

$$Q_1\ ,\ Q_2^2+Q_3^2\ ,\ Q_3^3 - 3Q_3Q_2^2\ . \tag{30}$$

A further condition that must be imposed on the three-body term is that it must go to zero if any one of the three atoms goes to infinity. The following general family of functions can be chosen for the three-body part to conform to the functional form adopted for the two-body part:

$$U_{ijk}^{(3)}/D = P(Q_1,Q_2,Q_3).F(a_3,Q_1), \tag{31}$$

where $P(Q_1,Q_2,Q_3)$ is a polynomial in the Q coordinates and F is a damping function, containing a single parameter, $a_3$, which determines the range of the three-body potential. Three different kinds of damping functions can be adopted:

$$\begin{aligned} F(a_3,Q_1) &= exp(- a_3Q_1) && \text{exponential ,} \\ F(a_3,Q_1) &= (½)[1- tanh(a_3Q_1/2)] && \text{tanh ,} \\ F(a_3,Q_1) &= sech(a_3Q_1) && \text{sech .} \end{aligned} \tag{32}$$

The use of the exponential damping function can lead to a problem, namely, for large negative $Q_1$ values (i.e. for triangles for which $r_{ij} + r_{jk} + r_{ki} << 3r_e$), the function $F$ may be large so that the three-body contribution swamps the total two-body contribution. This may lead to the collapse of the lattice. To overcome this problem, it may be necessary in some cases, to add a hard wall function to the repulsive part of the two-body term.

The polynomial, $P$, is normally taken to be

$$\begin{aligned} P(Q_1,Q_2,Q_3) = c_o + c_1Q_1 + c_2Q_1^2 + c_3(Q_2^2+Q_3^2 ) + c_4Q_1^3 \\ + c_5Q_1(Q_2^3+Q_3^2) + c_6(Q_3^3-3Q_3Q_2^2 ). \end{aligned} \tag{33}$$

This implies that there are seven parameters to be determined. For systems where simultaneous fitting is made to data for two different solid phases the following quartic terms can be added

$$C_7Q_1^4 + c_8Q_1^2.(Q_2^2+Q_3^2) + c_9(Q_2^2+Q_3^2)^2 + c_{10}Q_1(Q_3^3-3Q_3Q_2^2). \tag{34}$$

The potential parameters for a set of elements are given in Table II

| Element | $a_2$ | $a_3$ | $D$(eV) | $r_e$(nm) | $c_o$ | $c_1$ | $c_2$ |
|---|---|---|---|---|---|---|---|
| Al | 7.0 | 8.0 | 0.9073 | 0.27568 | 0.2525 | - 0.4671 | 4.4903 |
| Cu | 7.0 | 9.0 | 0.888 | 0.2448 | 0.202 | -0.111 | 4.990 |
| Ag | 7.0 | 9.0 | 0.722 | 0.2799 | 0.204 | -0.258 | 6.027 |
| Sn | 6.25 | 3.55 | 1.0 | 0.2805 | 1.579 | -0.872 | -4.980 |
| Pb | 8.0 | 6.0 | 0.59273 | 0.332011 | 0.18522 | 0.87185 | 1.27047 |

| Element | $c_2$ | $c_4$ | $c_5$ | $c_6$ | $c_7$ | $c_8$ | $c_9$ | $c_{10}$ |
|---|---|---|---|---|---|---|---|---|
| Al | -1.1717 | 1.6498 | -5.3579 | 1.6327 | 0.0 | 0.0 | 0.0 | 0.0 |
| Cu | -1.369 | 0.469 | -2.630 | 1.202 | 0.0 | 0.0 | 0.0 | 0.0 |
| Ag | -1.262 | -0.442 | -5.127 | 2.341 | 0.0 | 0.0 | 0.0 | 0.0 |
| Sn | -13.145 | -4.781 | 35.015 | -1.505 | 2.949 | -15.065 | 10.572 | 12.830 |
| Pb | -3.44145 | -3.884 | 15 5.27033 | 2.85596 | 0.0 | 0.0 | 0.0 | 0.0 |

*Table II: Parameters of the Murrell- Mottram Potentials*

### 3.1.5. The many-body Rafii-Tabar and Sutton (RTS) long-range alloy potentials

We now consider the case of many-body interatomic potentials that describe the energetics of metallic alloys, and in particular the FCC metallic alloys. The interatomic potential that models the energetics and dynamics of a binary, A-B, alloy is normally constructed from the potentials that separately describe the A-A and the B-B interactions, where A and B are the elemental metals. To proceed with this scheme, a combining rule is normally proposed. Such a rule would allow for the computation of the A-B interaction parameters from those of the A-A and B-B parameters. The combining rule reflects the different averaging procedures that can be adopted, such as the arithmetic or the geometric averaging. The criterion for choosing any one particular combining rule is the closeness of the results obtained, when computing with the proposed A-B potential obtained with that rule, with the corresponding experimental values where they exist.

The Rafii-Tabar and Sutton potentials [11, 58] are the generalization of the SC potentials and model the energetics of the metallic FCC random binary alloys. They have the advantage that all the parameters for the alloys are obtained from those for the elemental metals without the introduction of any new parameters. The basic form of the potential is given by

$$U^{RTS} = (½)\sum_i \sum_{j \neq i} \hat{p}_i \; \hat{p}_j \, V^{AA}(r_{ij}) + (1- \hat{p}_i)(1- \hat{p}_j) V^{BB}(r_{ij})$$

$$+ [\hat{p}_i (1-\hat{p}_j) + \hat{p}_j (1-\hat{p}_i)] V^{AB}(r_{ij})$$

$$- d^{AA} \sum_i \hat{p}_i [\sum_{j \neq i} \hat{p}_j \Phi^{AA}(r_{ij}) + (1-\hat{p}_j) \Phi^{AB}(r_{ij})]^{1/2}$$

$$- d^{BB} \sum_i (1-\hat{p}_i) [\sum_{j \neq i} (1-\hat{p}_j) \Phi^{BB}(r_{ij}) + \hat{p}_j \Phi^{AB}(r_{ij})]^{1/2}. \quad (35)$$

The operator $\hat{p}_i$ is the site occupancy operator and is defined as

$\hat{p}_i = 1$ if site i is occupied by an A atom

$\hat{p}_i = 0$ if site i is occupied by a B atom (36)

The functions $V^{\alpha\beta}$ and $\Phi^{\alpha\beta}$ are defined as

$$V^{\alpha\beta}(r) = \varepsilon^{\alpha\beta} [a^{\alpha\beta} / r]^{n_{\alpha\beta}} ,$$

(37)

$$\Phi^{\alpha\beta}(r) = [a^{\alpha\beta} / r]^{m_{\alpha\beta}}$$

where $\varepsilon$ and $\beta$ are both A and B. The parameters $\varepsilon^{AA}$ , $c^{AA}$ , $a^{AA}$ , $m^{AA}$ and $n^{AA}$ are for the pure element A, and $\varepsilon^{BB}$ , $c^{BB}$ , $a^{BB}$, $m^{BB}$ and $n^{BB}$ are for the pure element B, given in Table I.

$$d^{AA} = \varepsilon^{AA} c^{AA} ,$$

$$d^{BB} = \varepsilon^{BB} c^{BB} . \quad (38)$$

The mixed, or alloy, states, are obtained from the pure states by assuming the combining rules:

$$V^{AB} = (V^{AA} V^{BB})^{1/2} , \quad (39)$$

$$\Phi^{AB} = (\Phi^{AA} \Phi^{BB})^{1/2} . \quad (40)$$

These combining rules, based on purely empirical grounds, give the alloy parameters as

$$m^{AB} = 1/2\ (m^{AA} + m^{BB})\ ,$$

$$n^{AB} = 1/2\ (n^{AA} + n^{BB})\ ,$$

$$a^{AB} = (a^{AA} a^{BB})^{½}\ ,$$

$$\varepsilon^{AB} = (\varepsilon^{AA} \varepsilon^{BB})^{½}\ . \qquad (41)$$

These potentials were used to compute the elastic constants and heat of formation of a set of FCC metallic alloys [58], as well as to model the formation of ultra thin Pd films on Cu(100) surface [59]. They form the basis of a large class of MD simulations [11, 33].

### 3.1.6. Angular-dependent potentials

Transition metals form three rather long rows in the Periodic Table, beginning with Ti, Zr and Hf and terminating with Ni, Pd and Pt. These rows correspond to the filling of 3d, 4d and 5d orbital shells, respectively. Consequently, the d-band interactions play an important role in the energetics of these metals [77], giving rise to angular-dependent forces that contribute significantly to the structural and vibrational characteristics of these elements. Pseudopotential models are commonly used to represent the intermolecular interaction in such metals [78, 79]. Recently, an *ab initio* generalized pseudopotential theory [80] was employed to construct an analytic angular-dependent potential for the description of the element Mo [81], a BCC transition metal. According to this prescription, the total cohesive energy is expressed as

$$H_I^{MO} = H_{vol}(\Omega) + (½_N) \sum_i \sum_{j \neq i} V_2(ij)$$
$$+ (^1/_{6N}) \sum_i \sum_{j \neq i} \sum_{k \neq i,j} V_3(ijk)$$
$$+ (^1/_{24N}) \sum_i \sum_{j \neq i} \sum_{k \neq i,j} \sum_{l \neq i,j,k} V_4(ijkl) \qquad (42)$$

where $\Omega$ is the atomic volume, $N$ is number of ions, $V_3$ and $V_4$ are, respectively, the angular-dependent three- and four-ion potentials and $H_{vol}$ includes all one-ion intraatomic contributions to the cohesive energy. The interatomic potentials, $V_2(_{ij})$, $V_3(_{ijk})$ and $V_4(_{ijkl})$ denote

$$V_2(_{ij}) \equiv V_2(r_{ij};\ \Omega)\ ,$$

$$V_3(_{ijk}) \equiv V_3(r_{ij},\ r_{jk},\ r_{kl};\ \Omega)\ ,$$

$$V_4(_{ijkl}) \equiv V_4(r_{ij},\ r_{jk},\ r_{kl},\ r_{li},\ r_{ki},\ r_{li};\ \Omega)\ , \tag{43}$$

where $r_{ij}$, for example, is the ion-ion separation distance between ions i and j. These potentials are expressible in terms of weak pseudopotential and d-state tight-binding and hybridization matrix elements that couple different sites. Analytic expressions for these functions are provided [80, 81] in terms of distances and angles subtended by these distances.

The potential expressed by Eq. (42) was employed to compute the values of a set of physical properties of Mo including the elastic constants, the phonon frequencies and the vacancy formation energy [81]. These results clearly show that the inclusion of the angular-dependent potentials greatly improves the computed values of these properties as compared with the results obtained exclusively from an effective two-body interaction potential, $V_2^{eff}$ . Furthermore, the potential was employed in an MD simulation of the melting transition of the Mo, details of which can be found in [81].

## 3.2. Interatomic potentials for covalently-bonding systems

### 3.2.1. The Tersoff many-body C-C, Si-Si and C-Si potentials

The construction of Tersoff many-body potentials are based on the formalism of analytic bond-order potential, initially suggested by Abell [82]. According to Abell's prescription, the binding energy of an atomic many-body system can be computed in terms of pairwise nearest-neighbor interactions that are, however, modified by the local atomic environment. Tersoff employed this prescription to obtain the binding energy in Si [83-85], C [86], Si-C [85, 87], Ge and Si-Ge [87] solid-state structures.

In the Tersoff's model, the total binding energy is expressed as

$$H_I^{TR} = \sum_i E_i = (½)\sum_i \sum_{j \neq i} V(r_{ij}) \ , \tag{44}$$

where $E_i$ is the energy of site i and $V(r_{ij})$ is the interaction energy between atoms i and j, given by

$$V(r_{ij}) = f_c(r_{ij}) \, [V^R(r_{ij}) + b_{ij} V^A(r_{ij})] \ . \tag{45}$$

The function $V^R(r_{ij})$ represents the repulsive pairwise potential, such as the core-core interactions, and the function $V^A(r_{ij})$ represents the attractive bonding due to the valence electrons. The many-body feature of the potential is represented by the term $b_{ij}$ which acts as the bond-order term and which depends on the local atomic environment in which a particular bond is located. The analytic forms of these potentials are given by

$$V^R(r_{ij}) = A_{ij} \exp(- \lambda_{ij} \, r_{ij}),$$

$$V^A(r_{ij}) = - B_{ij} \exp(- \mu_{ij} \, r_{ij}),$$

$$f_c(r_{ij}) = 1, \qquad \text{for} \quad r_{ij} < R_{ij}^{(1)},$$

$$f_c(r_{ij}) = (½)+(½)\cos\left[\pi(r_{ij}-R_{ij}^{(1)})/(R_{ij}^{(2)}-R_{ij}^{(1)})\right], \qquad \text{for} \quad R_{ij}^{(1)} < r_{ij} < R_{ij}^{(2)},$$

$$f_c(r_{ij}) = 0, \qquad \text{for} \quad r_{ij} > R_{ij}^{(2)}$$

$$b_{ij} = \chi_{ij} [1 + (\beta_i \zeta_{ij})^{n_i}]^{-0.5 n_i} ,$$

$$\zeta_{ij} = \sum_{k \neq i,j} f_c(r_{ik}) \omega_{ik} \, g(\theta_{ijk}) ,$$

$$g(\theta_{ijk}) = 1 + c_i^2/d_i^2 - c_i^2 / [d_i^2 + (h_i - \cos\theta_{ijk})^2 ] ,$$

$$\lambda_{ij} = (\lambda_i + \lambda_j)/2 , \qquad \mu_{ij} = (\mu_i + \mu_j)/ 2 ,$$

$$\omega_{ik} = \exp[\mu_{ik}(r_{ij} - r_{ik})]^3 ,$$

$$A_{ij} = \sqrt{A_i A_j}, \qquad B_{ij} = \sqrt{B_i B_j} ,$$

$$R_{ij}^{(1)} = \sqrt{R_i^{(1)} R_j^{(1)}}, \qquad R_{ij}^{(2)} = \sqrt{R_i^{(2)} R_j^{(2)}}, \qquad (46)$$

Numerical values of the parameters of Tersoff potentials for C and Si are listed in Table III.

| Parameter | *C* | *Si* |
|---|---|---|
| A(ev) | $1.3936 \times 10^3$ | $1.8308 \times 10^3$ |
| B(ev) | $3.467 \times 10^2$ | $4.7118 \times 10^2$ |
| $\lambda$ ( $nm^{-1}$ ) | 34.879 | 24.799 |
| $\mu$ ( $nm^{-1}$ ) | 22.119 | 17.322 |
| $\beta$ | $1.5724 \times 10^{-7}$ | $1.1000 \times 10^{-6}$ |
| $\eta$ | $7.2751 \times 10^{-1}$ | $7.8734 \times 10^{-1}$ |
| *c* | $3.8049 \times 10^4$ | $1.0039 \times 10^5$ |
| *d* | 4.384 | 16.217 |
| *h* | -0.57058 | -0.59825 |
| $R^{(1)}$ (nm) | 0.18 | 0.27 |
| $R^{(2)}$ (nm) | 0.21 | 0.30 |
| $\chi$ | 1 | 1 |
| $\chi_{C\text{-}Si}$ | 0.9776 | |

*Table III: Parameters of the Tersoff potentials for C and Si*

where the labels i, j and k refer to the atoms in the ijk bonds, $r_{ij}$ and $r_{ik}$ refer to the lengths of the ij and ik bonds whose angle is $\theta_{ijk}$. Singly subscripted parameters, such as $\lambda_i$ and $n_i$, depend only on one type of atom, e.g. C or Si. The parameters for the C-C, Si-Si and Si-C potentials are listed in Table III. For the C, the parameters were obtained by fitting the cohesive energies of carbon polytypes, along with the lattice constant and bulk modulus of diamond. For the Si, the parameters were obtained by fitting to a database consisting of cohesive energies of real and hypothetical bulk polytypes of Si, along with the bulk modulus and bond length in the diamond structure. Furthermore, these potential parameters were required to reproduce all three elastic constants of Si to within 20%.

### 3.2.2. The Brenner-Tersoff type first generation hydrocarbon potentials

The Tersoff potentials correctly model the dynamics of a variety of solid-state structures, such as the surface reconstruction in Si [83, 84] or the formation of interstitial defects in carbon [86]. However, while these potentials can give a realistic description of the C-C single, double and triple bond lengths and energies in hydrocarbons, solid graphite and diamond, they lead to non-physical results for bonding situations intermediate between the single and double bonds, such as the bonding in the Kekul´e construction for the graphite where, due to bond conjugation, each bond is considered to be approximately one-third double-bond and two-thirds single-bond in character. To correct for this, and similar problems in hydrocarbons, as well as to correct for the non-physical overbinding of radicals, Brenner [88] developed a Tersoff-type potential for hydrocarbons that can model the bonding in a variety of small hydrocarbon molecules as well as in diamond and graphite. In this potential, Eq.s (44) and (45) are written as

$$H_I^{Br} = (½)\sum_i \sum_{i \neq j} V(r_{ij}) \qquad (47)$$

and

$$V(r_{ij}) = f_c(r_{ij}) \left[ V^R(r_{ij}) + \bar{b}_{ij} V^A(r_{ij}) \right], \qquad (48)$$

where

$$V^R(r_{ij}) = D_{ij}/(S_{ij}-1).\exp\left[-\sqrt{(2S_{ij})}.\beta_{ij}\,(r_{ij} - R_{ij}^{e})\right],$$

$$V^A(r_{ij}) = -\,D_{ij}S_{ij}/(S_{ij}-1).\exp\left[-\sqrt{(2S_{ij})}.\beta_{ij}\,(r_{ij} - R_{ij}^{e})\right],$$

$$\bar{b}_{ij} = (b_{ij} + b_{ji})/2 + F_{ij}\,(N_i^{(t)},N_j^{(t)},N_{ij}^{Conj}),$$

$$b_{ij} = [1 + G_{ij} + H_{ij}\,(N_i^{(H)},N_i^{(C)})]^{-\delta_i},$$

$$G_{ij} = \sum_{k\neq i,j} f_c(r_{ik})\,G_i(\theta_{ijk}).\exp\left[\alpha_{ijk}\{(r_{ij} - R_{ij}^{(e)}) - (r_{ik} - R_{ik}^{(e)})\}\right],$$

$$G_c(\theta) = a_o\,[1 + c_o^2/d_o^2 - c_o^2/[d_o^2 + (1+\cos\theta)^2]. \qquad (49)$$

The quantities $N_i^{(C)}$ and $N_i^{(H)}$ represent the number of C and H atoms bonded to atom i, $N_i^{(t)}=(N_i^{(C)}+N_i^{(H)})$ is the total number of neighbors of atom i and its values, for neighbors of the two carbon atoms involved in a bond, can be used to determine if the bond is part of a conjugated system. For example, if $N_i^{(t)}<4$, then the carbon atom forms a conjugated bond with its carbon neighbors. $N_{ij}^{conj}$ depends on whether an ij carbon bond is part of a conjugated system. These quantities are given by

$$N_i^{(H)} = \sum_{l\neq i,j}^{\text{hydrogen atoms}} f_c(r_{il}),$$

$$N_i^{(C)} = \sum_{k\neq i,j}^{\text{carbon atoms}} f_c(r_{ik}),$$

$$N_{ij}^{conj} = 1 + \sum_{k\neq i,j}^{\text{carbon atoms}} f_c(r_{ik})F(x_{ik}) + \sum_{l\neq i,j}^{\text{carbon atoms}} f_c(r_{jl})F(x_{jl}),$$

$$F(x_{ik}) = 1, \qquad \text{for} \qquad x_{ik}\leq 2$$

$F(x_{ik}) = ½+ (½)cos[\pi(x_{ik} - 2)],$ for $2<x_{ik}< 3$

$F(x_{ik}) = 0,$ for $x_{ik}\geq 3$

$$x_{ik} = N_k^{(t)} - f_c(r_{ik}). \quad (50)$$

The expression for $N_{ij}^{conj}$ yields a continuous value as the bonds break and form, and as the second-neighbor coordinations change. For $N_{ij}^{conj}=1$ the bond between a pair of carbon atoms i and j is not part of a conjugated system, whereas for $N^{conj}\geq 2$ the bond is part of a conjugated system.

The functions $H_{ij}$ and $F_{ij}$ are parameterized by two- and three- dimensional cubic splines respectively, and the potential parameters in Eqs. (47) to (50) were determined by first fitting to systems composed of carbon and hydrogen atoms only, and then the parameters were chosen for the mixed hydrocarbon systems. Two sets of parameters, consisting of 63 and 64 entries, are listed in [88]. These parameters were obtained by fitting a variety of hydrocarbon data sets, such as the binding energies and lattice constants of graphite, diamond, simple cubic and FCC structures, and the vacancy formation energies. The complete fitting sets are given in Tables I, II and III in [88].

### 3.2.3. The Brenner-Tersoff-type second generation hydrocarbon potentials

The potential function, expressed by Eqs. (47)-(50) and referred to as the first generation hydrocarbon potential, was recently further refined [41, 89] by including improved analytic functions for the intramolecular interactions, and by an extended fitting database, resulting in a significantly better description of bond lengths, energies and force constants for hydrocarbon molecules, as well as elastic properties, interstitial defect energies, and surface energies for diamond. In this improved version, the terms in Eq. (48) are redefined as

$$V^R(r_{ij}) = f_c(r_{ij}).[1 + Q_{ij}/r_{ij}]\, A_{ij}.exp(\alpha_{ij}\, r_{ij}),$$

$$V^A(r_{ij}) = - f_c(r_{ij}) \sum_{(n=1,3)} B_{ijn}.exp(\beta_{ijn}\, r_{ij}),$$

$$\bar{b}_{ij} = (p_{ij}^{\sigma\pi} + p_{ji}^{\sigma\pi})/ 2 + p_{ij}^{\pi} ,$$

$$p_{ij}^{\pi} = \pi_{ij}^{rc} + \pi_{ij}^{dh} ,$$

$$p_{ij}^{\sigma\pi} = \left[1 + G_{ij} + P_{ij}(N_i^{(H)}, N_i^{(C)}\right]^{-½},$$

$$G_{ij} = \sum_{k\neq i,j} f_c(r_{ik}) G_i\, [\cos(\theta_{jik})].\exp[\lambda_{ijk}(r_{ij} - r_{ik}) ,$$

$$\pi_{ij}^{rc} = F_{ij}\, (N_i^{(t)}, N_j^{(t)}, N_{ij}^{conj}),$$

$$N_{ij}^{conj} = 1 + [\sum_{k\neq i,j}^{\text{carbon atoms}} f_c(r_{ik})F(x_{ik})]^{2} + [\sum_{l\neq i,j}^{\text{carbon atoms}} f_c(r_{jl})F(x_{jl})]^{2},$$

$$\pi_{ij}^{dh} = T_{ij}\, (N_i^{(t)}, N_j^{(t)}, N_{ij}^{conj}).\left[\sum_{k\neq i,j}\sum_{l\neq i,j} (1 - \cos^2\omega_{ijkl}) f_c(r_{ik}).f_c(r_{jl})\right],$$

$$\cos\omega_{ijkl} = e_{ijk}.e_{ijl}\, . \qquad (51)$$

$Q_{ij}$ is the screened Coulomb potential, which goes to infinity as the interatomic distances approach zero. The term $\pi_{ij}^{rc}$ represents the influence of radical energetics and $\pi$-bond conjugation on the bond energies, and its value depends on whether a bond between atoms i and j has a radical character and is part of a conjugated system. The value of $\pi_{ij}^{dh}$ depends on the dihedral angle for the C-C double bonds. $P_{ij}$ represents a bicubic spline, $F_{ij}$ and $T_{ij}$ are tricubic spline functions. In the dihedral term, $\pi_{ij}^{dh}$, the functions $e_{jik}$ and $e_{ijl}$ are unit vectors in the direction of the cross products $\boldsymbol{R}_{ji}\times\boldsymbol{R}_{ik}$ and $\boldsymbol{R}_{ij}\times\boldsymbol{R}_{jl}$, respectively, where the $\boldsymbol{R}$'s are the interatomic vectors. The function $G_c[\cos(\theta_{jik})]$ modulates the contribution that each nearest-neighbor makes to $\bar{b}_{ij}$. This function was determined in the following way. It was computed for the selected values of $\theta=109.47^o$ and $\theta=120^o$, corresponding to the bond angles in diamond and graphitic sheets, and for $\theta = 90^o$ and $\theta = 180^o$, corresponding to the bond angles among the nearest neighbors in a simple cube lattice. The FCC lattice contains angles of $60^o$, $90^o$, $120^o$ and 180°. A value of $G_c[cos(\theta = 60^o)]$ was also computed from the above values. To complete an analytic function for the $G_c[cos(\theta)]$, sixth order polynomial splines in $cos(\theta)$ were used to obtain its values for $\theta$ between $109.47^o$ and $120^o$. For $\theta$ between $0^o$ and $109^o$, for a carbon atom i, the angular function

$$g_c = G_c[cos(\theta)] + Q(N_i^{(t)}).\left[\gamma_c cos(\theta) - G_c\{cos(\theta)\}\right], \qquad (52)$$

is employed, where $\gamma_c\, cos(\theta)$ is a second spline function, determined for angles less than $109.47^o$. The function $Q(N_i^{(t)})$ is defined by

$$\begin{aligned} Q(N_i^{(t)}) &= 1, && \text{for} \quad N_i^{(t)} \leq 3.2, \\ Q(N_i^{(t)}) &= ½+(½)cos[\pi(N_i^{(t)}-3.2)/(3.7-3.2)], && \text{for} \quad 3.2<N_i^{(t)}<3.7, \\ Q(N_i^{(t)}) &= 0, && \text{for} \quad N_i^{(t)} \geq 3.7 \end{aligned} \qquad (53)$$

The large database of the numerical data on parameters and spline functions were obtained by fitting the elastic constants, vacancy formation energies and the formation energies for interstitial defects for diamond.

## 3.3. Interatomic potential for C-C non-bonding systems

The non-bonding interactions between carbon atoms are required in many of the simulation studies in computational nanoscience and nanotechnology. These can be modeled according to various types of potentials. The Lennard-Jones and Kihara potentials can be employed to describe the van der Waals intermolecular interactions between carbon clusters, such as $C_{60}$ molecules, and between the basal planes in a graphite lattice. Other useful potentials are the exp-6 potential [90] which also describes the $C_{60}$-$C_{60}$ interactions, and the Ruoff-Hickman potential [91] which models the $C_{60}$-graphite interactions.

### 3.3.1. The Lennard-Jones and Kihara potentials

The total interaction potential between the carbon atoms in two $C_{60}$ molecules, or between those in two graphite basal planes, could be represented by the Lennard-Jones potential [92]

$$H_I^{LJ}(r_{ij}^{IJ}) = 4\varepsilon\sum_i\sum_{j>i} [(\sigma/r_{ij}^{IJ})^{12} - (\sigma/r_{ij}^{IJ})^6] , \qquad (54)$$

where I and J denote the two molecules (planes), $r_{ij}$ is the distance between the atom i in molecule (plane) i and atom j in molecule (plane) J. The parameters of this potential, *($\varepsilon$=0.24127 × $10^{-2}$ ev , $\sigma$=0.34 nm),* were taken from a study of graphite [93]. The Kihara potential is similar to the Lennard-Jones except for the fact that a third parameter d, is added to correspond to the hard-core diameter, i.e.

$$H_I^{LJ}(r_{ij}^{IJ}) = 4\varepsilon\sum_i\sum_{j>i} [\{(\sigma\text{-}d)/(r_{ij}^{IJ}\text{-}d)\}^{12} - [\{(\sigma\text{-}d)/(r_{ij}^{IJ}\text{-}d)\}^6] \quad \text{for } r>d,$$

$$H_I^{LJ}(r_{ij}^{IJ}) = \infty \qquad \text{for } r \le d \qquad (54\text{-}1)$$

### 3.3.2. The exp-6 potential

This is another potential that describes the interaction between the carbon atoms in two $C_{60}$ molecules

$$H_I^{EXP6}(r_{ij}^{IJ}) = \sum_i\sum_{j>i} [A \exp(- \alpha r_{ij}^{IJ})- B /(r_{ij}^{IJ})^6]. \qquad (55)$$

Two sets of values of the parameters are provided, and these are listed in Table IV. These parameters have been obtained from the gas phase data of a large number of organic compounds, without any adjustment. The measured value of the $C_{60}$ solid lattice constant is a = 1.404 nm at T = $11^o$ K. The calculated value using the set one was a =

1.301 nm and using the set two was a = 1.403 nm. The experimentally estimated heat of sublimation is equal to - 45kcal/mol (extrapolated from the measured value of - 40.1 ± 1.3 kcal/mol at T = 707$^o$ K). The computed value using the set one was - 41.5 kcal/mol and using the set two was - 58.7 kcal/mol. We see that whereas the set two produces a lattice constant nearer the experimental value, the thermal properties are better described by using the set one.

| | *A(kcal/mol)* | *B [kcal/mol × (nm)$^6$]* | $\alpha$ *(nm)$^{-1}$* |
|---|---|---|---|
| Set one | 42000 | 3.58 x $10^8$ | 35.8 |
| Set two | 83630 | 5.68 x $10^8$ | 36.0 |

*Table IV: Parameters of the exp-6 potential for C.*

### 3.3.3. The Ruoff-Hickman potential

This potential, based on the model adopted by Girifalco [94], describes the interaction of a $C_{60}$ molecule with a graphite substrate by approximating these two systems as continuum surfaces on which the carbon atoms are 'smeared out' with a uniform density. The sums over the pair interactions are then replaced by integrals that can be evaluated analytically. The $C_{60}$ is modeled as a hollow sphere having a radius b = 0.355 nm, and the C-C pair interaction takes on a Lennard-Jones form

$$H_l(r_{ij}) = c_{12} r^{-12} - c_6 r^{-6}, \tag{56}$$

with $c_6 = 1.997\times10^{-5}$ $[ev.(nm)^6]$ and $c_{12} = 3.4812\times10^{-8}$ $[ev.(nm)^{12}]$ [94]. The interaction potential between the hollow $C_{60}$ and a single carbon atom of a graphite substrate, located at a distance z > b from the center of the sphere, is then evaluated as

$$V(z) = V_{12}(z) - V_6(z)\ , \tag{57}$$

where

$$V_n(z) = c_n/[2(n-2)].[N/(bz)].[1/(z - b)^{n-2} - 1/(z + b)^{n-2}], \tag{58}$$

where $N$ is the number of atoms on the sphere ($N = 60$ in this case) and $n = 12, 6$. The total interaction energy between the $C_{60}$ and the graphite plane is then obtained by integrating $V(z)$ over all the atoms in the plane, giving

$$H_l(R) = E_{12}(R) - E_6(R), \qquad (59)$$

where

$$E_n(R) = \{c_n/[4(n-2)(n-3)]\}.(N^2/b^3).\left[1/(R-b)^{n-3} - 1/(R+b)^{n-3}\right], \qquad (60)$$

and $R$ is the vertical distance of the center of the sphere from the plane.

## 3.4 Interatomic potential for metal-carbon system

In modeling the growth of metallic films on semi-metallic substrates, such as graphite, a significant role is played by the interface metal-carbon potential since it controls the initial wetting of the substrate by the impinging atoms and also determines the subsequent diffusion and the final alignments of these atoms. This potential has not been available and we have used an approximate scheme, based on a combining rule, to derive its general analytic form [95]. To construct a mixed potential to describe the interaction of an FCC metallic atom (M) with C, we assumed a generalized Morse-like potential energy function

$$H_l^{MC}(r_{ij}) = \sum_i\sum_{j>i} E_{MC}\left[exp\{-N\alpha(r_{ij} - r_w)\} - N.exp\{-\alpha(r_{ij} - r_w)\}\right], \qquad (61)$$

and to obtain its parameters, we employed a known Morse potential function

$$H_l^{CC}(r_{ij}) = \sum_i\sum_{j>i} E_C \left[exp\{- 2\alpha_1(r_{ij} - r_d)\} - 2exp\{- \alpha_1(r_{ij} - r_d)\}\right], \qquad (62)$$

that describes the C-C interactions [96], and a generalized Morse-like potential function

$$H_l^{MM}(r_{ij}) = \sum_i\sum_{j>i} E_M \left[\exp\{- m\alpha_2(r_{ij} - r_o)\} - m.\exp\{- \alpha_2(r_{ij} - r_o)\}\right], \tag{63}$$

that describes the M-M interactions [97]. Several combining rules were then tried. The rule giving the satisfactory simulation results led to

$$E_{MC} = \sqrt{E_C\, E_M}\ ,$$

$$r_w = \sqrt{r_d\, r_o}\ ,$$

$$\alpha = \sqrt{\alpha_1\, \alpha_2},$$

$$N = \sqrt{2m}. \tag{64}$$

Since a cut-off is normally applied to an interaction potential, the zero of this potential at a cut-off, $r_c$, was obtained according to the prescription in [96] leading to

$$H_l^{MC}(r_{ij}) = \sum_i\sum_{j>i} E_{MC} \left[\exp\{- N\, \alpha\, ( r_{ij} - r_w)\} - N\exp\{- \alpha( r_{ij} - r_w)\}\right] - E_{MC} \left[\exp\{- N\alpha( r_c - r_w)\} - N\exp\{- \alpha( r_c - r_w)\}\right]$$
$$- E_{MC}N\alpha/\eta\ \left[1 - \exp\, \eta( r_{ij} - r_c) \right] \times \left[\exp(- N\alpha( r_c - r_w)) - \exp( - \alpha( r_c - r_w)) \right], \tag{65}$$

where $\eta$ is a constant whose value was chosen to be $\eta$ =20. This was a sufficiently large value so that the potential (84) was only modified near the cut-off distance. The parameters, pertinent to the case when the metal atoms were silver, i.e M=Ag, are listed in Table V. The parameters for Eq. (62) were obtained by fitting the experimental cohesive energy and the inter-planar spacing, c /2 , of the graphite exactly, and the parameters for Eq. (63) were obtained by fitting the experimental values of the stress-free lattice parameter and elastic constants $C_{11}$ and $C_{12}$ of the metal.

| | |
|---|---|
| $\alpha_1$ | 49.519 $(nm)^{-1}$ |
| $\alpha_2$ | 3.7152 $(nm)^{-1}$ |
| $E_C$ | 3.1 ev |
| $E_{Ag}$ | 0.0284875 ev |
| $M$ | 6.00 |
| $r_o$ | 0.444476 nm |
| $r_d$ | 0.12419 nm |

*Table V: Parameters of the Ag-C potential.*

## 3.5. Atomic-site stress field

In many modeling studies involving the mechanical behavior of nanostructures, such as the simulation of the dynamics of crack propagation in an atomic lattice, it is necessary to compute a map of the stress distribution over the individual atomic sites in a system composed of N atoms.

The concept of atomic-level stress field was developed by Born and Huang [98] using the method of small homogeneous deformations. Applying small displacements to a pair of atoms i and j, with an initial separation of $r_{ij}$, it can be shown that [99] the Cartesian components of the stress tensor at the site i are given by

$$\sigma_{\alpha\beta}(i) = (1/2\Omega_i)\sum_{j>i}[\partial\Phi(r_{ij})/\partial r_{ij}].[r_{ij}^{\alpha}r_{ij}^{\beta}/r_{ij}], \qquad (66)$$

where $\alpha, \beta = x, y, z$, $\Phi(r_{ij})$ is the two-body central potential, and $\Omega_i$ is the local atomic volume which can be identified with the volume of the Voronoi polyhedron associated with the atom i [100].

For the many-body potential energy given by Eq. (35), the stress tensor is given by

$$\sigma_{\alpha\beta}^{RTS}(i) = (1/2\Omega_i)\Big[\sum_{j\neq i}[\partial V(r_{ij})/\partial r_{ij}]$$
$$- (1/2)d^{AA}\,\hat{p}_i \sum_{j\neq i}(1/\sqrt{\rho_i^A} + 1/\sqrt{\rho_j^A}).[\partial\Phi^A(r_{ij})/\partial r_{ij}]$$
$$- (1/2)d^{BB}(1-\hat{p}_i)\sum_{j\neq i}(1/\sqrt{\rho_i^B} + 1/\sqrt{\rho_j^B}).[\partial\Phi^B(r_{ij})/\partial r_{ij}]\Big].(r_{ij}^{\alpha}r_{ij}^{\beta})/r_{ij}, \qquad (67)$$

which for an elemental lattice with the two-body potentials given in [37] reduces to (see also [101])

$$\sigma_{\alpha\beta}^{RTS}(i) = (\varepsilon/a^2)(1/2\Omega_i)\Big[\sum_{j\neq i} [- n(a/r_{ij})^{n+2} + cm(1/\sqrt{\rho_i} + 1/\sqrt{\rho_j})(a/r_{ij})^{m+2}\Big] (r_{ij}^{\alpha} r_{ij}^{\beta}), \quad (68)$$

where only the contribution of the virial component to the stress field has been included and the contribution of the kinetic energy part (momentum flux) has been ignored as we are only interested in the low-temperature stress distributions. The volumes associated with individual atoms, $\Omega_i$, can be obtained by computing numerically their corresponding Voronoi polyhedra according to the prescription given in [23].

## 3.6. Direct measurement of interparticle forces by atomic force microscope (AFM)

The invention of the atomic force microscope, AFM, [7] in 1986 and its modification to optical detection [102] has opened new perspectives for various micro- and nanoscale surface imaging in science and industry. The use of AFM not only allows for nanoscale manipulation of the morphology of various condensed phases and the determination of their electronic structures, it can be also used for direct determination of interatomic and intermolecular forces.

However, its use for measurement of interparticle interaction energies as a function of distance is getting more attention due to various reasons. For atoms and molecules consisting of up to ten atoms , quantum mechanical *ab initio* computations are successful in producing rather exact force-distance results for interparticle potential energy. For complex molecules and macromolecules one may produce the needed intermolecular potential energy functions directly only through the use of atomic force microscope (AFM). For example, atomic force microscopy data are often used to develop accurate potential models to describe the intermolecular interactions in the condensed phases of such molecules as $C_{60}$ [103].

The atomic force microscope (AFM) is a unique tool for direct study of intermolecular forces. Unlike traditional microscopes, AFM does not use optical lenses and therefore it provides very high-resolution range of various sample properties [7,104,105]. It operates by scanning a very sharp tip across a sample, which 'feels' the contours of the surface in a manner similar to the stylus tracing across the grooves of a record. In this way it can follow the contours of the surface and so create a topographic image, often with sub-nanometer resolution.

This instrument also allows researchers to obtain information about the specific forces between and within molecules on the surface. The AFM, by its very nature, is extremely sensitive to intermolecular forces and has the ability to measure force as a function of distance. In fact measurement of interactions as small as a single hydrogen bond have been reported [106-110]. The non-contact AFM will be used for attractive interactions force measurement. Contact AFM will be used for repulsive force measurement. Intermittent-contact AFM is more effective than non-contact AFM for imaging larger scan sizes.

In principle to do such a measurement and study with AFM it is necessary to specially design the tip for this purpose [102,111,112]. Sarid [8] has proposed force-distance relationships when the tip is made of a molecule, a sphere, and a cylinder assuming van der Waals dispersion attractive forces. Various other investigators have developed the methodologies for force-distance relationship for other tip geometric shapes including cylinder, paraboloid, cone, pyramid, a conical part covered by the spherical cap, etc [105, 111, 113-119]. For example, Zanette et al [111] present a theoretical and experimental investigation of the force–distance relation in the case of a pyramidal tip. Data analysis of interaction forces measured with the atomic force microscope is quite important [120]. Experimental recordings of direct tip–sample interaction can be obtained as described in [121] and recordings using flexible cross-linkers can be obtained as described in [122,123]. The noise in the typical force–distance cycles can be assumed to be, for example, Gaussian.

Recent progress in AFM technology will allow the force-distance relationship measurement of inter- and intra-molecular forces at the level of individual molecules of almost any size.

Because of the possibility to use the AFM in liquid environments [109, 124] it has become possible to image organic micelles, colloids, biological surfaces such as cells and protein layers and generally organic nanostructures [4] at nm-resolution under physiological conditions. One important precaution to be considered in the force measurement is how to fix micelles, colloids, biological cells on a substrate and a probe, securely enough for measuring force but flexible enough to keep the organic nanostructure intact and in case of biological cells keep it biologically active [124]. Variety of techniques for this purpose have been proposed including the use of chemical cross-linkers, flexible spacer molecules [125], inactive proteins as cushions in case of biological systems [126] and self-assembled monolayers [127]. An important issue to consider in liquid state force-distance measurements is the effect of pushing the organic nanostructures on the substrate and AFM probe. As the AFM probe is pushed onto the nanostructure, there is a possibility of damaging it or adsorbing it to the probe physically.

Also making microelectrophoretic measurements of zeta potential will allow us to calculate the total interparticle energies indirectly. From the combined AFM and microelectrophoretic measurements accurate force-distance data can be obtained. From the relation between the force and distance, an interparticle force vs. distance curve can be created. Then with the use of the phenomenological potential functions presented in this review the produced data can be readily fitted to a potential energy function for application in various nanotechnology and nanoscience computational schemes.

## 3.7. Conclusions and Recommendations:

In this review we have presented a set of state-of-the-art phenomenological interatomic and intermolecular potential energy functions that are widely used in computational modeling at the nanoscale. We have also presented a review of direct measurement of interparticle force-distance relationship from which intermolecular potential energy functions data can be generated. There is still a great deal of work need to be doe in order to develop a thorough database for interatomic and intermolecular potential energy functions to be sufficient for applications in nanoscience and naotechnology. This is because to control the matter atom by atom, molecule by molecule and/or at the macromolecular level, which is the aim of the nanotechnology, it is necessary to know the exact intermolecular forces between the particles under consideration. In the development of intermolecular force models applicable for the study of nanostructures which are at the confluence of the smallest of human-made devices and the largest molecules of living systems it is necessary to reexamine the existing techniques and come up with more appropriate intermolecular force models.

It is understood that formidable challenges remain in the fundamental understanding of various phenomena in nanoscale before the potential of nanotechnology becomes a reality. With the knowledge of better and more exact intermolecular interactions between atoms and molecules it will become possible to increase our fundamental understanding of nanostructures. This will allow development of more controlable processes in nanotechnology and optimization of production and design of more appropriate nanostructures, like nanotubes [128] and its interactions with other nanosystems.